\documentclass[conference]{IEEEtran}
\IEEEoverridecommandlockouts
\usepackage{cite}
\usepackage{amsmath,amssymb,amsfonts,mathtools, bm}
\usepackage{amsthm}
\usepackage{algorithm}
\usepackage{algorithmicx}
\usepackage{algpseudocode}
\usepackage{graphicx}
\usepackage{textcomp}
\usepackage{xcolor,float}
\usepackage{tabularx}
\usepackage{textcomp}
\usepackage{diagbox}
\usepackage{booktabs}
\usepackage{subfigure}
\newcounter{MYtempeqncnt}

\allowdisplaybreaks
\renewcommand{\baselinestretch}{1}

\begin{document}

\title{Label-free Deep Learning Driven Secure Access Selection in Space-Air-Ground Integrated Networks}
\author{
\IEEEauthorblockN{
Zhaowei Wang\IEEEauthorrefmark{1},
Zhisheng Yin\IEEEauthorrefmark{2},
Xiucheng Wang\IEEEauthorrefmark{2},
Nan Cheng\IEEEauthorrefmark{2},
Yuan Zhang\IEEEauthorrefmark{3},
and Tom H. Luan\IEEEauthorrefmark{1}
}
\IEEEauthorblockA{
\IEEEauthorrefmark{1}School of Cyber Engineering, Xidian University, Xi’an, 710071, China\\
\IEEEauthorrefmark{2}School of Telecommunications Engineering, Xidian University, Xi’an, 710071, China\\
\IEEEauthorrefmark{3}University of Electronic Science and Technology of China, Chengdu 611731, China\\
Email: 1916040218@s.upc.edu.cn; \{zsyin, tom.luan\}@xidian.edu.cn; xcwang\_1@stu.xidian.edu.cn; \\dr.nan.cheng@ieee.org; zy\_loye@126.com}}
    \maketitle

\IEEEdisplaynontitleabstractindextext

\IEEEpeerreviewmaketitle

\begin{abstract}
In Space-air-ground integrated networks (SAGIN), the inherent openness and extensive broadcast coverage expose these networks to significant eavesdropping threats. Considering the inherent co-channel interference due to spectrum sharing among multi-tier access networks in SAGIN, it can be leveraged to assist the physical layer security among heterogeneous transmissions. However, it is challenging to conduct a secrecy-oriented access strategy due to both heterogeneous resources and different eavesdropping models. In this paper, we explore secure access selection for a scenario involving multi-mode users capable of accessing satellites, unmanned aerial vehicles, or base stations in the presence of eavesdroppers. Particularly, we propose a Q-network approximation based deep learning approach for selecting the optimal access strategy for maximizing the sum secrecy rate. Meanwhile, the power optimization is also carried out by an unsupervised learning approach to improve the secrecy performance. Remarkably, two neural networks are trained by unsupervised learning and Q-network approximation which are both label-free methods without knowing the optimal solution as labels. Numerical results verify the efficiency of our proposed power optimization approach and access strategy, leading to enhanced secure transmission performance.
\end{abstract}

\begin{IEEEkeywords}
Space-air-ground integrated network, label-free deep learning, access selection, secrecy rate

\end{IEEEkeywords}

\section{Introduction}

Space-air-ground integrated network (SAGIN), as a novel network design that integrates satellites, aerial platforms, and ground communication systems, integrates heterogeneous networks to enhance network resource utilization and wireless access transmission capacity, attracting wide attention in academia and industry \cite{CHENG20221}. The heterogeneity, self-organization, and time-varying characteristics of SAGIN offer significant advantages for various services and applications, but also pose challenges such as resource allocation and management, power control, and secure transmission \cite{9390169,inbook}. In particular, due to the open nature and broad coverage of wireless channels, as well as the dynamic topology of satellite networks, communication in SAGIN is vulnerable to eavesdropping threats. Moreover, the complex and extensive geographical environment provides ample hiding space for potential attackers and eavesdroppers (Eves), resulting in serious security issues \cite{9797229}. Ensuring secure transmission in SAGIN networks has become an urgent problem to be addressed.

In the context of SAGIN, multi-mode terminals are devices that can seamlessly connect to satellite, unmanned aerial vehicles (UAVs), base stations (BSs), and other heterogeneous networks. They can autonomously select the most secure mode of access for communication based on the perceived channel characteristics. The selection strategy of terminals is influenced by the randomness of wireless channels, and the access strategies of different multi-mode terminal devices can mutually affect each other. Therefore, determining the most secure communication access method has become a key consideration for multi-mode terminals.

Physical layer security, which utilizes the characteristics of wireless channels and employs physical layer techniques for secure communication, can serve as an effective complement to traditional encryption techniques based on key systems, providing comprehensive information security protection \cite{9700672}. Physical layer security leverages the differences in randomness between channels to achieve lightweight secure transmission. When the quality of the main channel is better than that of the eavesdropping channel, legitimate users can achieve secure transmission through lossless encoding. However, in scenarios with a large number of antennas or users, traditional physical layer security optimization algorithms can become very complex \cite{8904319}.

Artificial intelligence has brought new opportunities to the research of physical layer security with its vigorous development. Deep learning, as a significant branch of artificial intelligence, demonstrates excellent performance in handling large-scale data. Its main objective is to construct models and extract features from sample data by learning the underlying patterns, which are suitable for predicting the future and making decisions based on current data. And after training, the inference computation of deep learning is much simpler than traditional algorithms, as it can be called multiple times after training, which can save computational resources. Currently, numerous experts and scholars have proposed solutions to physical layer security issues based on deep learning, and achieved promising results \cite{wang2022joint,9020300,article,9860451,9078075}. However, there is limited literature that addresses the security access selection problem of multi-mode terminals in heterogeneous networks using deep learning, which motivates the focus of this work. Moreover, traditional deep learning methods depend on optimal solutions as training labels to achieve high performance, which are challenges to obtain in the security access selection problem. Therefore, it is necessary to design label-free methods to reduce the training cost of neural network-based methods\cite{wang2023scalable}.

In this paper, we investigate the secure access selection in SAGIN, where multi-mode terminal users are considered and the spectrum sharing is executed among downlinks of satellite, UAV, and BS. To guarantee secure transmissions and realize intelligent secure access,
we formulate a problem to maximize the sum secrecy rate of multi-mode users, where their lowest communication rates are also guaranteed, respectively, and downlink transmission powers are constrained. To address such an intractable optimization problem, a joint framework is proposed where the optimal access selection strategy is obtained by a Q-network approximation-based deep learning approach, and the power allocation of corresponding access points is solved by an unsupervised learning approach. Both two neural networks are trained label-free without knowing optimal solution as labels. Finally, the secrecy performance of our proposed approach is evaluated and its efficiency is verifed.

\section{System Model}

\begin{figure}[ht]
  \centering
  \includegraphics[width=0.85\columnwidth]{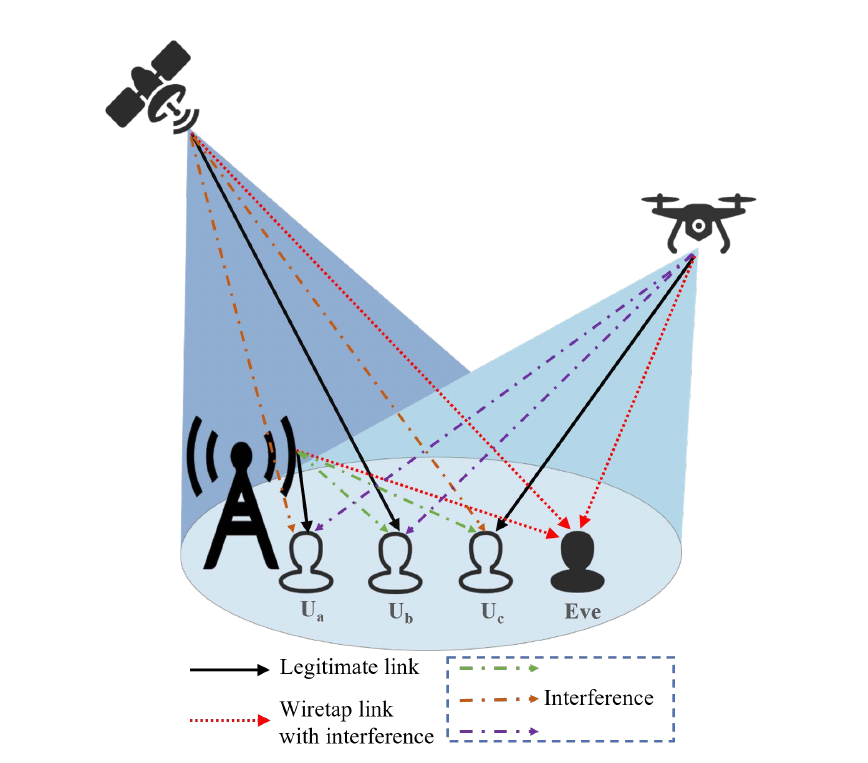}
   \vspace{-9pt}
  \centering \caption{Downlink transmission for multi-modal users in SAGIN} 
  \label{system}
   \vspace{-9pt}
\end{figure}

Fig. \ref{system} depicts a diagram of access and transmission scenario for multi-mode users in presence of an Eve in downlink of SAGIN, where the multi-mode user can access to satellite, UAV, and BS arbitrarily. Particularly, we consider three legtimiate users,  e.g., User A ($\text{U}_a$), user B ($\text{U}_b$), and user C ($\text{U}_c$) coexisting within the overlapping coverage of satellite, UAV and BS. 
A worst-case eavesdropping scenario is considered in which Eve is located within that overlapping area and operates on the same frequency band as these users. Thus, either of $\text{U}_a$, $\text{U}_b$, and $\text{U}_c$ could be eavesdropped by the Eve. 
Furthermore, spectrum sharing occurs among the downlinks of satellite, UAV, and BS, resulting in potential co-channel interference. Nevertheless, with appropriate design, this interference can be treated as green interference to enhance the security of the system. Fig. \ref{system} only presents a common scenario of access.

\subsection{Physical Layer Channel Models}
In the access scenario of multi-mode users in SAGIN, there are three types of communication links, including satellite-to-ground link, air-to-ground link, and ground link.

The formula for the satellite-to-ground link is given by \cite{9645156}:
\begin{align}
    &h=\sqrt{C_Lb\beta}\exp \left( -j\theta \right), 
\end{align}    
where $C_L$ represents free space loss, and its formula is given as $C_L=\left( \lambda /4\pi \right) ^2/\left( d^2+l^2 \right)$, Where $\lambda$ represents the wavelength of the signal, $l$ represents the horizontal distance from the satellite beam center to the ground user, and $l$ represents the height of the satellite. $\beta$ represents the channel gain caused by rain attenuation, which follows a log-normal random variable, i.e., $\ln \left( \beta _{dB} \right) \thicksim \mathcal{N}\left( \mu ,\delta ^2 \right)$ with $\beta _{dB}$ is the dB form of $\beta$. $\theta$ is a phase vector uniformly distributed in the range $\left[ 0,2\pi \right)$. $b$ represents the satellite beam gain, which is defined as:
\begin{align}
    &b=G\left( \frac{J_1\left( u_0 \right)}{2u_0}-36\frac{J_3\left( u_0 \right)}{u_{0}^{2}} \right) ^2,
\end{align}  
where $G$ represents the maximum gain of the satellite antenna, $u_0=2.07123\sin \left( \alpha \right) /\sin \left( \alpha _{3dB} \right) $, $\alpha$ is the elevation angle between the beam center and the user, $\alpha _{3dB}$ is the 3dB angle of the satellite beam, $J_1\left( \cdot \right)$ and $J_3\left( \cdot \right)$ are the first and third order Bessel functions of the first kind. Therefore, assuming $h_{S,a}$, $h_{S,b}$, $h_{S,c}$, $h_{S,e}$ are the channel state information(CSI) from the satellite to $\text{U}_a$, $\text{U}_b$, $\text{U}_c$, and Eve, respectively.

The air-to-ground link can be defined as
\begin{align}
    &a=\sqrt{G_L}\left( \sqrt{\frac{K}{K+1}}a_{LoS}+\sqrt{\frac{1}{K+1}}a_{Ray} \right),
\end{align} 
where $G_L$ represents the path loss, given by $G_L=g_0/\left( U_{d}^{2}+U_{h}^{2} \right)$, $g_0$ represents the channel power gain at a reference distance of 1 m, $U_d$ is the horizontal distance from UAV to the target user, and $U_h$ is the height of the UAV. Small-scale fading follows the Rician channel model, where $K$ is the Rician factor, $a_{LoS}$ represents the line-of-sight Rician fading component, and $a_{Ray}$ represents the non-line-of-sight Rayleigh fading component. Therefore, assuming $h_{U,a}$, $h_{U,b}$, $h_{U,c}$, $h_{U,e}$ represent CSI from the UAV to $\text{U}_a$, $\text{U}_b$, $\text{U}_c$, and Eve, respectively.

The ground link can be defined as
\begin{align}
&g=\sqrt{\alpha}g_0,
\end{align} 
The parameter $\alpha$ represents large-scale fading, with the formula $\alpha =C_0r^{-4}$, $C_0$ denotes the channel power gain at a reference distance of 1 m, while $r$ represents the distance between the BS and the user. $g_0$ represents small-scale fading, following a Nakagami-$m$ distribution. Therefore, assuming $h_{B,a}$, $h_{B,b}$, $h_{B,c}$, $h_{B,e}$ represent CSI from the BS to $\text{U}_a$, $\text{U}_b$, $\text{U}_c$, and Eve, respectively.

\subsection{Transmission and Eavesdropping Model}
By defining the set of access points and the user index set, i.e., $\mathcal{A}= \left\{ {Sat,{UAV},BS} \right\} $ and $\mathcal{U}= \left\{ {a,b,c} \right\} $, respectively, 
$x_{i,u}\in \left\{ 0,1 \right\} $ denotes whether the user $u\in \mathcal{U}$ selects accessing $i \in \mathcal{A}$ or not, and $\sum_{i \in {\cal A}}{x_{i,u}}=1$ ensures that only  one access link is established for a same user at the same time. However, a case is involved that multiple users can be associated with the same access network.

We calculate the achievable rate of $\text{U}_a$, which can be expressed as

\begin{align*}
{R_{u_{a}}} = \sum\limits_{i \in \mathcal{A}} {{x_{i,a}}{{\log }_2}\left( {1 + \frac{{{p_{i,a}}{\mid h_{i,a}\mid ^2}}}{{\sum\limits_{u \in \mathcal{U}, u \ne a, j \in \mathcal{A}} {{p_{j,u}}{\mid h_{j,a}\mid ^2}}  + \delta _{i,a}^2}}} \right)},\tag{5}
\end{align*}
where $p_{i,a}$ denotes the power allocation to $\text{U}_a$ from its access point $i\in\mathcal{A}$, $h_{i,a}$ and $h_{j,a}$ denote the CSI from access point $i$ or $j$ to $\text{U}_a$, $\delta _{i,a}^2$ denotes the noise power received at $\text{U}_a$. Similarly, the achievable rate of $\text{U}_b$ and $\text{U}_c$ can be respectively obtained as
\begin{align*}
{R_{u_{b}}} = \sum\limits_{i \in {\cal A}} {{x_{i,b}}{{\log }_2}\left( {1 + \frac{{{p_{i,b}}{\mid h_{i,b}\mid ^2}}}{{\sum\limits_{u \in {\cal U},u \ne b,j \in {\cal A}} {{p_{j,u}}{\mid h_{j,b}\mid ^2}}  + \delta _{i,b}^2}}} \right)},\\
\tag{6} 
\end{align*}
\begin{align*}
{R_{u_{c}}} = \sum\limits_{i \in {\cal A}} {{x_{i,c}}{{\log }_2}\left( {1 + \frac{{{p_{i,c}}{\mid h_{i,c}\mid ^2}}}{{\sum\limits_{u \in {\cal U},u \ne c,j \in {\cal A}} {{p_{j,u}}{\mid h_{j,c}\mid ^2}}  + \delta _{i,c}^2}}} \right)}.\\
\tag{7}
\end{align*}

For the Eve, it receives a overlapping signal and has possibility to targeting either of $\text{U}_a$, $\text{U}_b$ and $\text{U}_c$. Thus the corresponding eavesdropping rate can be respectively written as
\begin{align*}
R_{e_{a}}=\sum\limits_{i\in \mathcal{A}}{x_{i,a}\log _2\left( 1+\frac{p_{i,a} \mid h_{i,e}\mid ^2}{\sum\limits_{u\in \mathcal{U},u\ne a,j\in \mathcal{A}}{p_{j,u}\mid h_{j,e} \mid ^2}+\delta _{e}^{2}} \right)},\\
\tag{8}
\end{align*}
\begin{align*}
R_{e_{b}}=\sum\limits_{i\in \mathcal{A}}{x_{i,b}\log _2\left( 1+\frac{p_{i,b}\mid h_{i,e} \mid ^2}{\sum\limits_{u\in \mathcal{U},u\ne b,j\in \mathcal{A}}{p_{j,u}\mid h_{j,e}\mid ^2}+\delta _{e}^{2}} \right)},\\
\tag{9}
\end{align*}
\begin{align*}
R_{e_{c}}=\sum\limits_{i\in \mathcal{A}}{x_{i,c}\log _2\left( 1+\frac{p_{i,c} \mid h_{i,e}\mid ^2}{\sum\limits_{u\in \mathcal{U},u\ne c,j\in \mathcal{A}}{p_{j,u}\mid h_{j,e} \mid ^2}+\delta _{e}^{2}} \right)},\tag{10}
\end{align*}
where $h_{i,e}$ and $h_{j,e}$ denote the CSI from access point $i$ or $j$ to Eve, $\delta _{e}^2$ denotes the noise power received at Eve.
For easy analysis, let $\delta _{i,a}^{2}=\delta _{i,b}^{2}=\delta _{i,c}^{2}= \delta_{e} ^2=1$.


According to the information-theoretic security, the secrecy rate is defined by
\begin{align*}
&R_s=\left[ C_u-C_e \right] ^+,\tag{11}
\end{align*}
where $C_u$ and $C_e$ denote the main and eavesdropping channel capacity, respectively.
Therefore,  the secrecy rates of $\text{U}_a$, $\text{U}_b$, and $\text{U}_c$ are respectively represented at the top of next page.

\begin{figure*}[ht]
\normalsize
\setcounter{MYtempeqncnt}{\value{equation}}
\begin{subequations}
\begin{align*}
R_{a}=&\left[\sum_{i\in \mathcal{A}}x_{i,a}\left( \log _{2}\left( 1+\frac{p_{i,a}\mid h_{i,a} \mid ^2}{\sum\limits_{u\in \mathcal{U},u\ne a,j\in\mathcal{A}}{p_{j,u}\mid h_{j,a}\mid ^2}+1} \right) -\log _{2}\left( 1+\frac{p_{i,a}\mid h_{i,e} \mid ^2}{\sum\limits_{u\in \mathcal{U},u\ne a,j\in \mathcal{A}}{p_{j,u}\mid h_{j,e} \mid ^2}+1}\right ) \right )\right]^+ ,\tag{12} 
\end{align*}
\end{subequations}

\begin{subequations}
\begin{align*}
R_b=&\left[\sum_{i\in \mathcal{A}}x_{i,b}\left( \log _2\left( 1+\frac{p_{i,b} \mid h_{i,b}\mid ^2}{\sum\limits_{u\in \mathcal{U},u\ne b,j\in \mathcal{A}}{p_{j,u}\mid h_{j,b}\mid ^2}+1} \right)-\log _2\left( 1+\frac{p_{i,b} \mid h_{i,e} \mid ^2}{\sum\limits_{u\in \mathcal{U},u\ne b,j\in \mathcal{A}}{p_{j,u}\mid h_{j,e} \mid ^2}+1} \right) \right)\right]^+, \tag{13}
\end{align*}
\end{subequations}

\begin{subequations}
\begin{align*}
R_c=&\left[\sum_{i\in \mathcal{A}}x_{i,c}\left( \log _2\left( 1+\frac{p_{i,c} \mid h_{i,c}\mid ^2}{\sum\limits_{u\in \mathcal{U},u\ne c,j\in \mathcal{A}}{p_{j,u}\mid h_{j,c} \mid ^2}+1} \right)-\log _2\left( 1+\frac{p_{i,c}\mid h_{i,e}\mid ^2}{\sum\limits_{u\in \mathcal{U},u\ne c,j\in \mathcal{A}}{p_{j,u}\mid h_{j,e}\mid^2}+1} \right) \right)\right]^+. \tag{14}
\end{align*}
\end{subequations}
\setcounter{equation}{\value{MYtempeqncnt}}
\hrulefill
\vspace*{4pt}
\end{figure*}


\subsection{Problem Formulation}
From (12-14), it can be observed that the co-channel interference impacts the signal-to-interference-plus-noise (SINR) of legitimate users and Eve, thus the secrecy rate performance could be leveraged by the strength of co-channel interference. Considering the variation of CSI associated with accessing link and adjustable power allocation for legitimate users, the secrecy rate performance will along with the access strategy.  
In order to improve the overall secrecy performance of multi-mode users access to SAGIN, we formulate a problem to maximize the sum secrecy rate of $\text{U}_a$, $\text{U}_b$, and $\text{U}_c$ as follows, where both access strategy and power allocation are addressed.
\begin{align*}
&\max_{\bm{p,x}} \sum\limits_{u \in \mathcal{U}} {{R_u}} \label{pro1}   \tag{15} \\
&s.t.\;\;\;R_{u_{a}}\ge Q_{\min},\tag{15a} \\
&\quad\quad\, R_{u_b}\ge Q_{\min},\tag{15b}\\
&\quad\quad\, R_{u_c}\ge Q_{\min},\tag{15c}\\
&\qquad \sum_{u\in \mathcal{U}}{x_{i,u}}p_{i,u}\le P_S,i\rightarrow Sat,\tag{15d} \\
&\qquad \sum_{u\in \mathcal{U}}{x_{i,u}}p_{i,u}\le P_U,i\rightarrow UAV,\tag{15e} \\
&\qquad \sum_{u\in \mathcal{U}}{x_{i,u}}p_{i,u}\le P_B,i\rightarrow BS,\tag{15f} \\
&\qquad \sum_{i\in \mathcal{A}}{x_{i,u}}=1,\tag{15g}\\
&\qquad x_{i,u}\in \left\{ 0,1 \right\},\tag{15h}
\end{align*}
where ${\bm{x}} = \left\{ {{x_{i,u}},i \in \mathcal{A},u \in \mathcal{U}} \right\}$ is associated with access selection strategy and ${\bm{p}} = \left\{ {{p_{i,u}},i \in \mathcal{A},u \in \mathcal{U}} \right\}$ is the vector variable of power allocation; (15a)-(15c) respectively constrain the communication rate of legitimate users which guarantee the common quality of service; (15d)-(15f) represent the transmission power constraints of satellite, UAV, and BS, respectively; (15g)-(15h) represent the access selection between the current access point and the user. It is worth noting that at a specific time, a user can only access to one access point, but an access point can associate with multiple users simultaneously. 

\addtolength{\topmargin}{0.1in}

 \section{Label-free Deep Learning for Power Optimization and Access Selection}
\begin{figure}[ht]
  \centering
  \includegraphics[width=0.85\columnwidth]{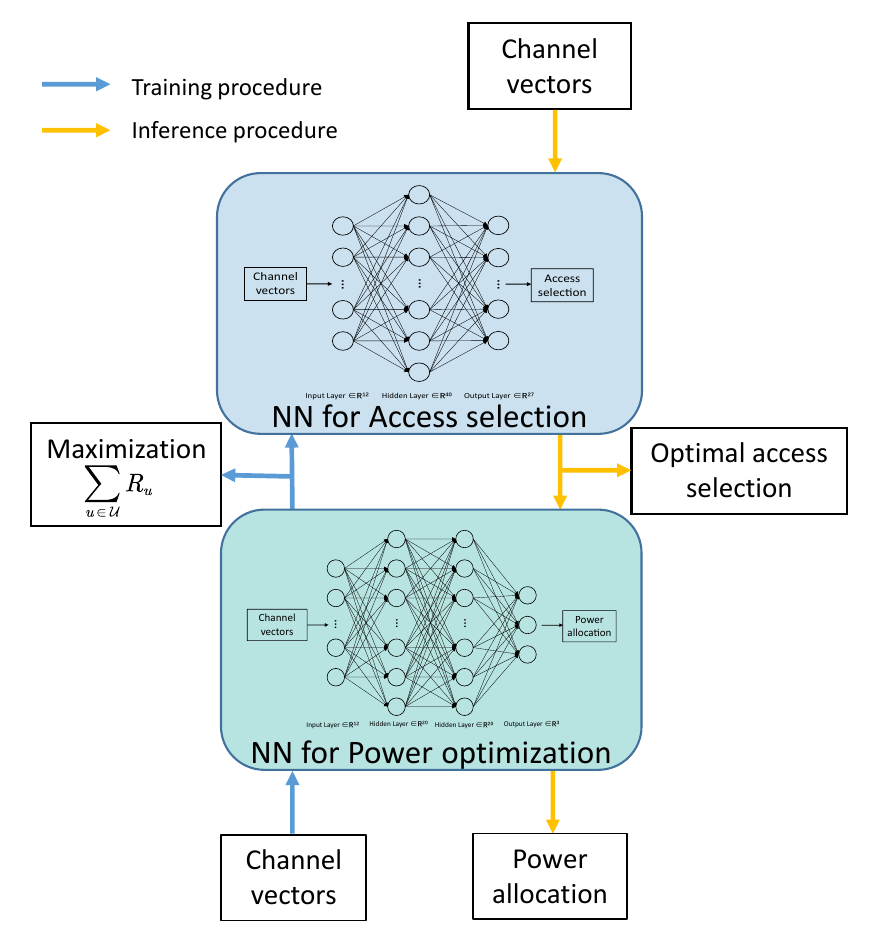}
   \vspace{-9pt}
  \centering \caption{Algorithm flowchart} 
  \label{network}
    \vspace{3mm}
   \vspace{-9pt}
\end{figure}

In this section, we propose a secure access strategy and power allocation methodology to address the non-convex problem formulated in (\ref{pro1}). To tackle this challenge, we divide the original problem into two subproblems and design a two-stage neural network (NN) architecture, as illustrated in Fig. \ref{network}. This architecture comprises a Q-network approximation-based deep learning approach for access selection and an unsupervised learning technique for power optimization. Notably, the label-free training and inference procedures are cascaded in Fig. \ref{network}.
\subsection{Power Optimization}

We employ unsupervised deep learning for power optimization, utilizing a three-layer multilayer perceptron (MLP) for training within each NN for power optimization. Both the first and second hidden layers comprise 20 neurons, mathematically represented as $N_1 = N_2 = 20$. Dropout layers, with a dropout rate of $\alpha = 0.2$, are incorporated after the hidden layers to mitigate overfitting. ReLU serves as the activation function after each layer, denoted as $\phi(\cdot) = \max(0, \cdot)$.

Throughout the training process, the parameters and weights of the NN for power optimization are continuously fine-tuned using the stochastic gradient descent (SGD) algorithm, causing the objective function $\sum\limits_{u \in \mathcal{U}} {{R_u}}$ to converge steadily until optimal secrecy performance is achieved. The power allocation for each user, corresponding to the NN for power optimization output, can be acquired.


The input vector $\bm{h}$ of the NN for power optimization  represent the CSI between the access points and access users, as well as between the access points and eavesdroppers, i.e., $\boldsymbol{h}=\left[ h_{S,a},h_{S,b},h_{S,c},h_{B,a},h_{B,b},h_{B,c},h_{U,a},h_{U,b},h_{U,c}, h_{S,e},h_{B,e},\right. $
$\left. h_{U,e}\right] $. The neural network optimizes power allocation by outputting a vector $\bm{p}^*$ representing the optimized power allocation values. As the objective of the neural network is to maximize the sum secrecy rate, the parameters $\bm{\theta}_{p}$ of the NN for power optimization $\mathcal{G}$ are updated according to the gradient update rule, which is defined as
\begin{align*}
    &\bm{\theta}_{p}= \bm{\theta}_{p} + lr_{p}\nabla_{\mathbf{p}} \mathcal{L}(\bm{\theta}_{p})|_{\bm{p}=\mathcal{G}(\bm{h,x})}\nabla_{\bm{\theta}_p}\mathcal{G}(\bm{h,x}),\tag{16}\label{gra}
\end{align*}
where $lr_p$ is the learning rate, $\mathcal{L}(\bm{\theta}_{p})$ is the loss function which is defined as
\begin{align*}
&\mathcal{L}(\bm{\theta}_{p})=-\varphi \sum\limits_{u \in \mathcal{U}} {{R_u}}+ \sum\limits_{k \in \mathcal{U}} \lambda_k \max \left( -R_{u_k}+Q_{{min}}, 0\right)            \\
&\quad\quad\;\, +\sum_{i\in \mathcal{A}}{\lambda _i \max \left( \sum_{u\in \mathcal{U}}{x_{i,u}p_{i,u}}-P_i, 0 \right)},\tag{17}\label{gra1}
\end{align*}
where $\lambda_i$, $\lambda_k$ and $\varphi$ represent adjustable parameters.

\subsection{Access Selection}
The primary objective of access selection training is to ascertain the access selection that yields superior security performance when given channel vectors. As the number of access points and access users significantly increases, compared to the approach of performing multiple power optimizations and comparing to obtain the maximum sum secrecy rate, making access selection decisions can reduce system running time, enhance system scalability, conserve computational resources, and decrease system latency.

We employed Q-network approximation for access selection training, with the neural network's objective being to approximate the Q-network. Training was conducted using a two-layer MLP. The hidden layer of the neural network contained 40 neurons, represented as $N_h = 40$, and dropout layers were utilized to prevent overfitting, with a dropout parameter set to $\alpha = 0.2$.


The set of $\sum\limits_{u \in \mathcal{U}} {{R_u}}$, which can be computed from the outputs of all power optimization networks' power values $\bm{p}^*$, is denoted as  $\boldsymbol{R}_{\boldsymbol{u}}\in R^{N\times 1}$. $\boldsymbol{h}$ also is the input of the NN for access selection $\mathcal{V}$. $\mathcal{V}\left( \boldsymbol{h} \right) \in R^{N\times 1}$ represents the output of the network. The loss function can be expressed as follows:
\begin{align*}
    \mathcal{L}\left( \boldsymbol{\theta }_{\mathcal{V}} \right) =\boldsymbol{\phi }\left(\left| \boldsymbol{R}_{\boldsymbol{u}}-\mathcal{V}\left( \boldsymbol{h}  \right) \right| \right), \tag{18}
\end{align*}
where $\boldsymbol{\phi }$ is a set of adjustable parameters, denoted as: $\boldsymbol{\phi }=\left[ \phi _1,\phi _2,\cdots ,\phi _N \right] $. N is the number of power optimization networks.

As the objective of the neural network is to approximate $\sum\limits_{u \in \mathcal{U}} {{R_u}}$, the parameters $\bm{\theta}_{\mathcal{V}}$ of the NN for access selection are updated using Adam algorithm according to the gradient update rule, which is defined as 
\begin{align*}
    &\bm{\theta} _{\mathcal{V}}=\bm{\theta} _{\mathcal{V}}-lr_{\mathcal{V}}\left(\boldsymbol{R}_{\boldsymbol{u}}-\mathcal{V}\left( \boldsymbol{h} \right) \nabla _{\bm{\theta} _{\mathcal{V}}}\mathcal{V}\left( \boldsymbol{h} \right) \right) ,\tag{19}\label{q}
\end{align*}
where $lr_{\mathcal{V}}$ is the learning rate.

Upon acquiring the two aforementioned training models, the inference process proceeds as follows: given an input channel vector, the access selection network model is initially employed to determine the current access choice. Subsequently, the power optimization network model corresponding to this access selection is utilized to derive the optimal power allocation, maximizing security performance under the current access conditions. The access selection and power allocation algorithm is concisely outlined in Algorithm 1.
\begin{algorithm}\label{A1}
\caption{Access Selection and Power Allocation}
\begin{algorithmic}[1]
\State \textbf{Training procedure:}
\State \textbf{Power Optimization:}
\For{each epoch}
    \State Input $\boldsymbol{h}$ for corresponding NN for power optimization;
    \State Obtain the corresponding power allocation;
    \State Calculate the loss function by equation (17) and Update the NN's parameters by equation (16) using SGD.
\EndFor
\State \textbf{Output}: The power allocation $\bm{p}^*$ for each access option.
\State \textbf{Access Selection:}
\For{each epoch}
    \State Use $\boldsymbol{h}$ as input for the NN for access selection;
    \State Calculate the loss by equation (18).
    \State Update the NN for access selection by equation (19) using the Adam and the Q-network approximation.
\EndFor
\State \textbf{Output}: The optimal access selection strategy.
\State \textbf{Inference procedure}:
\State \,  1) Input $\boldsymbol{h}$ to NN for access selection;
\State \,  2) NN for access selection output the optimal access strategy with the highest secrecy performance;
\State \;  3) Choose the corresponding NN for power optimization based on access strategy;
\State \;  4) Input the same $\boldsymbol{h}$ to NN for power optimization;
\State \;  5) NN for power optimization output the power allocation for maximizing security performance.
\end{algorithmic}
\end{algorithm}
\section{Numerical Results}

In this section, we establish a sophisticated SAGIN simulation platform and conduct a series of simulations to assess the sum secrecy rate performance of multi-mode users. For the satellite-to-ground link, parameters include a 600 km orbit altitude, a maximum beam gain of 52 dB, and a $0.4^{\circ}$ 3 dB beamwidth for the satellite beam. Rain attenuation is modeled using a log-normal random distribution, $\mathcal{N}\left( -3.125,1.6 \right)$. Horizontal distances between the satellite and $\text{U}_a$, $\text{U}_b$, $\text{U}_c$, and Eve are set at 2200 m, 2000 m, 2250 m, and 2250 m, respectively.
The ground link incorporates a channel power gain of -38.46 dB at a reference distance of 1 m, and distances between the BS and $\text{U}_a$, $\text{U}_b$, $\text{U}_c$, and Eve are 250 m, 120 m, 250 m, and 200 m, respectively. The Nakagami-$m$ fading model is employed with $m=2$ and an average power of $\varOmega =1$.
The air-to-ground link features a UAV altitude of 120 m, a channel power gain of -40 dB at a reference distance of 1 m, and a Rice factor $K$ of 10 dB. The carrier frequency for the satellite, UAV, and BS downlink is set at 2 GHz. Distances between the UAV and $\text{U}_a$, $\text{U}_b$, $\text{U}_c$, and Eve are 300 m, 260 m, 100 m, and 120 m, respectively. Through this comprehensive simulation framework, we aim to optimize secure communication strategies for heterogeneous networks and the secrecy rate performance is evaluated.
 \begin{figure}[ht]
  \centering
  \includegraphics[width=0.95\columnwidth]{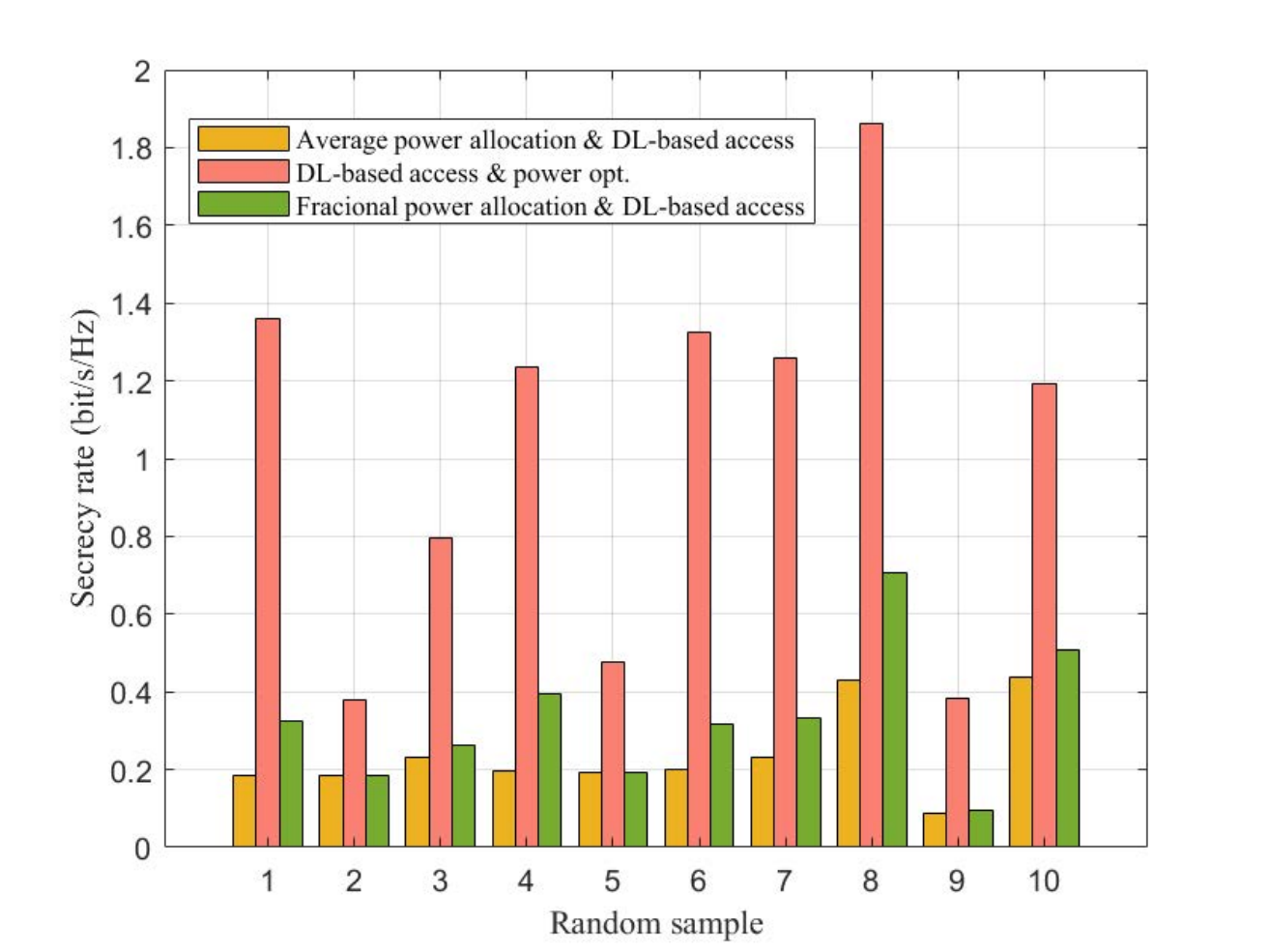}
   \vspace{-10pt}
  \centering \caption{The impact of power optimization on the sum secrecy rate. ($P_{S}$ = 12 dB, $P_B$ = 20 dB, $P_U$ = 3 dB,  $Q_{min}$ = 0.1 bit/s/Hz))} 
  \vspace{3mm}
  \label{Rsa}
   \vspace{-9pt}
\end{figure}

Fig. \ref{Rsa} illustrates the impact of power optimization on the secrecy rate performance, in which two fixed power allocation schemes are employed for comparison: the equal power allocation scheme, where the power is equally distributed among users, and the fractional power allocation scheme, which has a fixed power ratio set according to the individual CSI of each user. As can be seen from Fig. \ref{Rsa}, the proposed deep learning-based power optimization approach significantly improves the sum secrecy rate for multi-mode users secure access.
 \begin{figure}[ht]
  \centering
  \includegraphics[width=0.95\columnwidth]{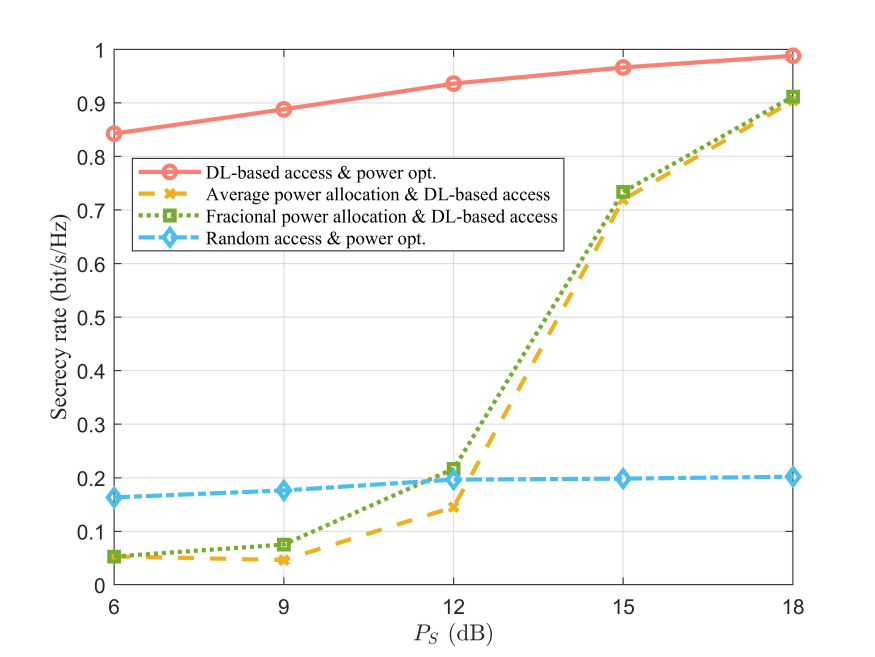}
   \vspace{-10pt}
  \centering \caption{The impact of the maximum transmission power of satellite on the sum secrecy rate. ($P_B$ = 15 dB, $P_U$ = 3 dB, $Q_{min}$ = 0.1 bit/s/Hz)} 
  \vspace{3mm}
  \label{PS}
   \vspace{-9pt}
\end{figure}

 \begin{figure}[ht]
  \includegraphics[width=0.95\columnwidth]{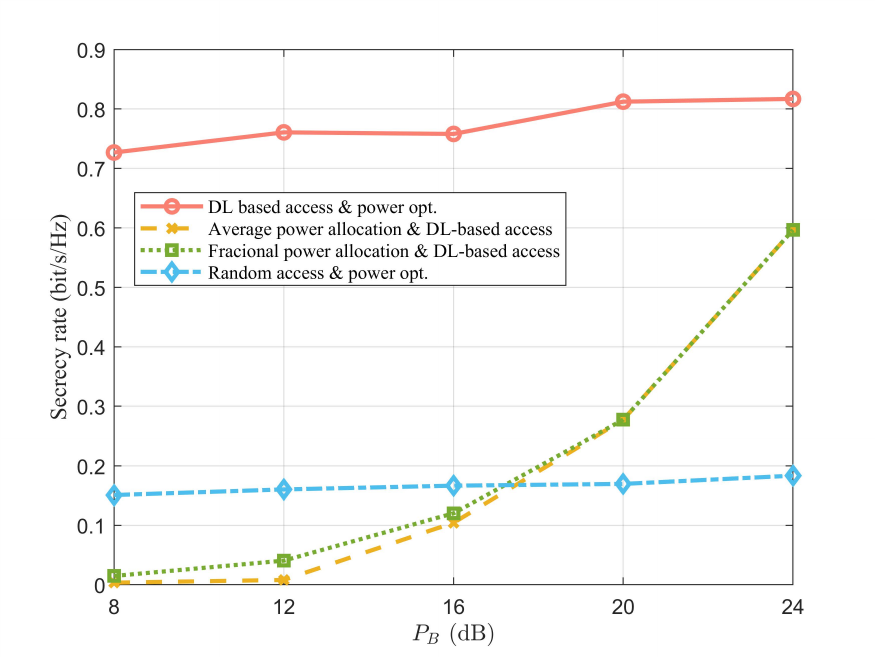}
   \vspace{-10pt}
  \centering \caption{The impact of the maximum transmission power of BS on the sum secrecy rate. ($P_S$ = 8 dB, $P_U$ = 6 dB, $Q_{min}$ = 0.1 bit/s/Hz)}  
  \vspace{3mm}
  \label{Pu}
   \vspace{-9pt}
\end{figure}

Fig. \ref{PS} and Fig. \ref{Pu} illustrate the impact of both transmission power of satellite and BS on the sum secrecy rate performance. From Fig. \ref{PS}, the approach proposed deep learning approach significantly enhances the sum secrecy rate for multi-mode users, comparing to fixed power allocation and random access schemes. Besides, the sum secrecy rate increases as satellite transmission power increases. This improvement is due to the growing capacity of the satellite link to influence the communication capacity of the access link and interfere with eavesdropping links as the satellite transmission power rises, thereby providing more advantageous access selection opportunities. In Fig. \ref{Pu}, the efficiency and improvement of our proposed approach compared to benchmarks are also verified.  

\section{Conclusion}
This paper has investigated the secure access problem of multi-mode users in SAGIN, where constraints of maximum transmission power and predefined communication rate are satisfied. To maximize the sum secrecy rate of multi-mode users, a novel approach is proposed that utilizes a cascaded Q-network approximation deep learning network and an unsupervised learning network to obtain the secure access selection strategy and the transmission power of downlinks, respectively. Particularly, these two NNs are trained by label-free approaches, which brings benefit of low complexity. In addition, simulations are carried out to verify the efficiency of our proposed algorithm in terms of secrecy performance, and an intelligent secure access scheme in multi-tier heterogeneous networks can be realized. 

\section*{Acknowledgement}
This work was supported in part by the National Natural Science Foundation of China (No. 62201432, 62071356, and 62101429), the Fundamental Research Funds for the Central Universities of Ministry of Education of China under Grant XJS221501, the National Natural Science Foundation of Shaanxi Province under Grant 2022JQ-602, and in part by the Guangzhou Science and Technology Program under Grant 202201011732.

\bibliography{ref}
\bibliographystyle{IEEEtran}

\ifCLASSOPTIONcaptionsoff
  \newpage
\fi

\end{document}